%% file: bare_conf.tex
\documentclass[conference]{IEEEtran}
\ifCLASSINFOpdf
\else
\fi

\usepackage{amsfonts,amsmath,amscd, amssymb}
\usepackage{graphicx}

\begin{document}
%
\title{Big Data Analytics for QoS Prediction Through Probabilistic Model Checking}

\author{\IEEEauthorblockN{Giuseppe Cicotti, Luigi Coppolino, Salvatore D'Antonio, Luigi Romano}
\IEEEauthorblockA{University of Naples, Parthenope\\
Centro Direzionale di Napoli,\\
80143 Naples - Italy \\
Email: \{giuseppe.cicotti,luigi.coppolino,salvatore.dantonio,luigi.romano\}@uniparthenope.it}
}


%


\maketitle

\begin{abstract}
As competitiveness increases, being able to guaranting QoS of delivered services is key for business success. It is thus of paramount importance the ability to continuously monitor the workflow providing a service and to timely recognize breaches in the agreed QoS level. The ideal condition would be the possibility to anticipate, thus predict, a breach and operate to avoid it, or at least to mitigate its effects. In this paper we propose a model checking based approach to predict QoS of a formally described process. The continous model checking is enabled by the usage of a parametrized model of the monitored system, where the actual value of parameters is continously evaluated and updated by means of big data tools.
The paper also describes a prototype implementation of the approach and shows its usage in a case study.


\end{abstract}


%
\IEEEpeerreviewmaketitle

\input{author}

%
%


\section*{Acknowledgment}
This work has been partially supported by the TENACE PRIN Project (n. 20103P34XC) funded by the Italian Ministry of Education, University and Research.
This work has been partially supported by the Embedded Systems in Critical Domains.
This project has received funding from the European Union's Seventh Framework Programme for research, technological development and demonstration under grant agreement no 313034 (SAWSOC Project).



%
%
%

\input{references}

\end{document}

%% file: author.tex
\section{Introduction}
\label{sec:intro}
The service-oriented computing paradigm has been changing the way of creating and developing software-based services. This paradigm is the foundation of the Utility Computing in which both hardware resources and software functionalities are made available according to the as-a-Service (aaS) model \cite{TSAI05}. 
This model allows developing new services by integrating and reusing existing ones, i.e. third party, or legacy systems. 
The result of such an integration is services being provided by extremily complex workflows \cite{WFR,BON}. Multiple parties are thus accountable for the successuful delivery of the service. The terms of service promised to the end-user of the provided service are described by means of an agreement normally named Service Level Agreement (SLA)\cite{BPM,SLA}. The terms of service regulating the relationship among parties collaborating to deliver the final service are named Operational Level Agreement (OLA). Both SLAs and OLAs are ultimatelly describing a QoS level to be matched while providing the service. 
Independentrly of the origin of QoS terms (SLA or OLA), it is of paramount importance the ability to continuously monitor the workflow providing the service and to timely recognize breaches in the agreed QoS levels \cite{QOS,QOS2}. The ideal condition would be the possibility to anticipate, thus predict, a breach and operate to avoid it, or at least to mitigate its effects.
In this paper we propose a new QoS prediction approach which combines run-time monitoring with a model-based analysis method such as probabilistic model-checking.
In our approach we use probabilistic model-checking to analyze a probabilistic model of the monitored workflow. To limit the state explotion problem, characteristic of model-checking analysis, we assume that the system analysed is described as a parametric model. At every moment of the analysis the actual value of parameters is retrieved, via run-time monitoring of the real-system, and it is used with the parametric model. The actualized model is used by the model checker which estimates the probability that in the very next future, the system will reach a status corresponding to an SLA violation. 
Since the amount of data retrieved during system monitoring can raise up very quickly \cite{ZZL13}, we use big data analytics solutions to guarantee the real-time evaluation of model parameters. \\
To validate the proposed approach we have developed a prototype of the QoS prediction framework and we have demonstrated its usage against a case study in the field of Smart Grids \cite{SG1,SG2}.

The paper is organized as follows. Section \ref{sec:relwork} describes related work on QoS prediction. In section \ref{sec:arch} we present the overall architecture enabling our QoS prediction approach. Section \ref{sec:casestudy} illustrates the case study to which we applied the proposed methodology, while section \ref{sec:valid} describes the prototype developed to validate our approach. Section \ref{sec:conc} closes the paper with conclusions and future work.

\section{Related Work}
\label{sec:relwork}

\input{rw}

\section{Architectural Overview}
\label{sec:arch}
\begin{figure}[ht!]
\centering
\includegraphics[scale=0.28]{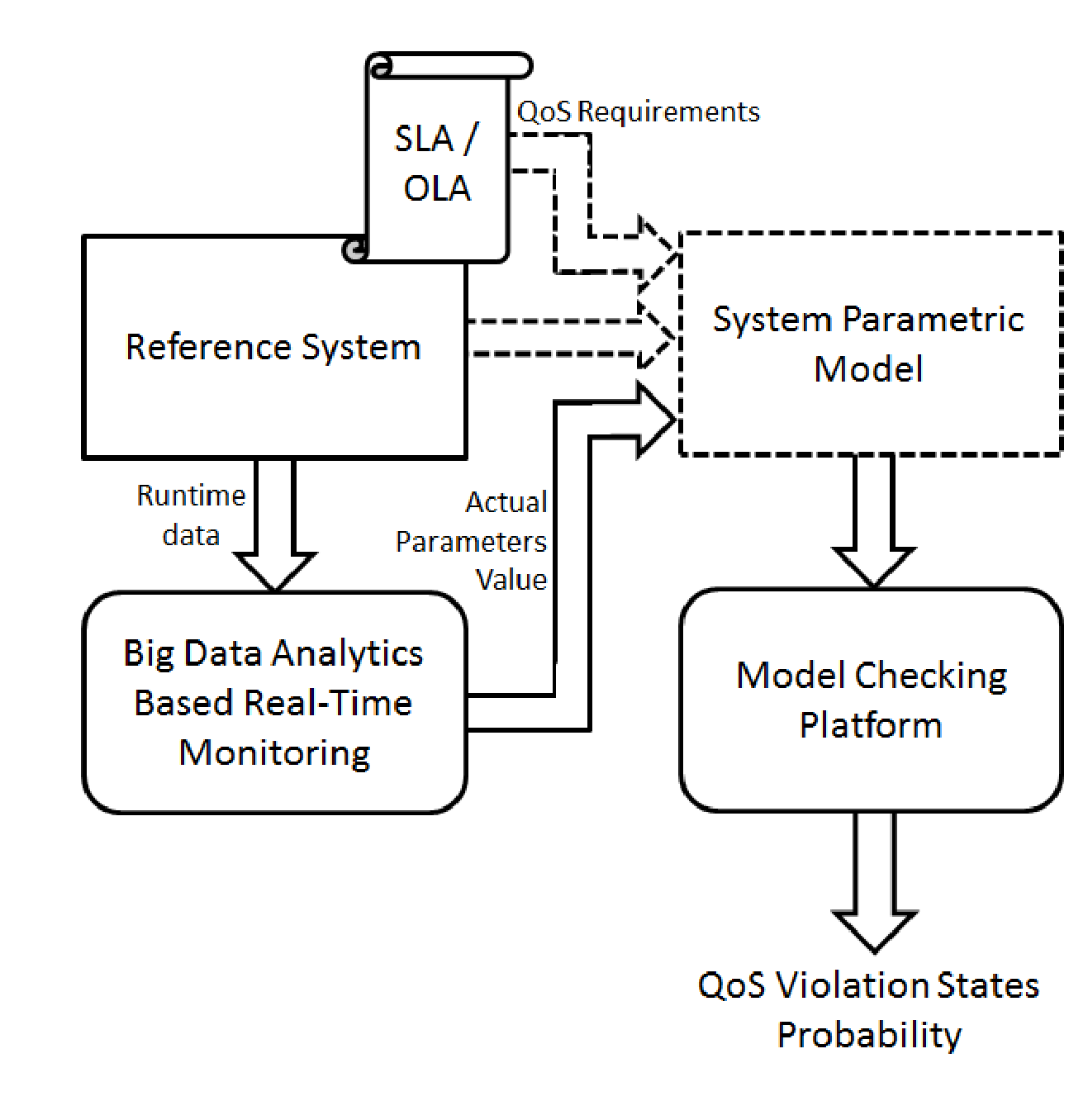}
\caption{The QoS Monitoring and Prediction achitecture}
\label{fig:arch}   
\end{figure}

This section describes the approach behind our solution for QoS prediction. An high level architecture overview is represented in fig. \ref{fig:arch}.

Given a system/process to be monitored for QoS compliance with a set of SLAs and OLAs, we assume that a formalized model of the system is made availabe.  Such a model is based on a state-transition description which is able to capture the evolution of Key Performance Indicators (KPIs) over time. Moreover states and transitions must be expressed as parameters. 
The KPIs can be inferred by the SLAs and OLAs defining the expected QoS. They can thus be used to identify the conditions of violation of the expected QoS. Such conditions are represented by some final states in the state-transition model. 

At run-time, the reference system is continously monitored and collected data are used to evaluate the actual value of model parameters. Once the model has been populated with estimated values of the parameters, it is processed by the model checking software. In our prototype we have used PRISM \cite{KNP11} which is an open-source probabilistic model checker which supports the analysis and checking of a wide number of model types.
The model checker explores states that can be reached since current state in a fixed number of transitions (depending on the the desired prediction time-lapse). If one of the states representing a violation is likely to be reached with a probability higher than a fixed threshold (violation alarm threshold), than a QoS breach is predicted. 

It is worth noting that the usage of a parametric model, which is continuously updated, and the fixed time-lapse used for the prediction, allow limiting the well-known state explosion problem due to the exhaustive states exploration operated by model-checkers. One further optimization could be operated by pruning those braches including states reachable with a probability lower than the violation alarm threshold.

As for the parameters evaluation, continous monitoring of a complex system may require the real-time analysis of huge amounts of data. Such requirement can be matched by using advaced Big Data Techniques and tools. In particular, in our prototype we used a Complex Event Processor (CEP) to infeer parameters value from collected data. In an advance prototype the Big Data layer could be used to support the model-checking process. 

To guarantee that the automatic procedure be both efficient and consistent, two conditions are to be held:

\begin{itemize}
\item the size of the state space of the model has to be sufficient to perform the model-checking analysis in a time that is compatible with the updating time of the QoS data of the modelled system
\item the evaluation of the model parameters should always allow the representation of the critical QoS states to be monitored.
\end{itemize} 

The first condition is key for obtaining a near real-time QoS prediction system. Indeed, it requests to balance the size of the QoS model at run-time by taking into account both the real time constraint imposed by the monitored service and the time spent to model check.
A preliminary analysis during the model definition has to be conduct in order to ensure that this condition is still true even though the model is fully expanded (i.e. no pruning of its state space is considered).
The second condition allows verifying that, if narrowed, the model still includes states of the real system related to critical QoS values (e.g. warning and/or violation states).

Consequently, our methodology considers the following steps:
\begin{enumerate}
\item Specification of the parameterised QoS stochastic model and QoS constraints to monitor
\item Real-time data analysis and parameters synthesis
\item Generation of the internal state-transition Model representation 
\item Execution of the Probabilistic Model-Checking to quantify the likelihood of future QoS state
\item QoS Verification
\end{enumerate}

In the first step we define a stochastic model which is suited to the kind of properties we are interested in monitoring. In this paper we show a case study, from the Smart-Grid domain, modelled by means of a CTMC.
The steps 2-5 are involved in an endless loop which makes our approach adaptive. 
In particular, the second step needs to analyse data received by the CEP so to determine the parameters of the model, and computes the current KPIs value. The third step generates the finite state-transition representation of the system model on which performing model-checking in the fourth step. Finally, the fifth step deals with verifying the QoS on the basis of the current KPIs and/or quantification of future QoS states.

\subsection{QoS Properties Specification}

In a previous work we introduced the concept of Quality Constraint (QC) \cite{CDC+13} as a mean to express constraints on KPIs. A QC is defined as a boolean condition on a single KPI. The language we used to specify QCs is an interval-based version of the Linear-time Temporal Logic (LTL). Particularly, in \cite{CDC+13} 
we introduced two temporal operators \textit{along} and \textit{within} which present the following semantic:
\begin{itemize}
\item $P$ \textbf{along} $T$: $P$ is true in any time instant belonging to $T$
\item $P$ \textbf{within} $T$: there exists at least a time instant $i \in T$ in which P is true
\end{itemize}

Thus, the \textbf{along} and \textbf{within} are, respectively, the restriction of the Linear Temporal Logic (LTL) ”globally” (G) and ”eventually” (F) operators to the interval $T$.
A QC without temporal operator is interpreted as an expression to be verified all along the lifetime of the monitored system, hence resulting useful for specifying safety property.

It is worth noting that in the context of runtime monitoring we check properties against execution traces of the system, i.e  ordered sequences of past (up to now) states. Although in this way we are able to recognise a violation as soon as it happens, we do not have any means to evaluate if it will happen and when in the future.

In this model-based approach we tackle this issue by defining Predictive Indicators (PIs) upon the monitored KPIs. A PI is a numerical indicator which statistically quantify the probability for a KPI to be in a certain state (i.e. a range of values) in a predetermined time instant in the future. Taking advantage of probabilistic model-checking we define such PIs as probabilistic temporal formulae (in the logic suitable for the underlying model) which can be evaluated over all possible evolution considered in the service model. Furthermore, as numerical indicators PIs can be monitored by means of specifying QCs.
To this purpose we have extended our QC language with the \textbf{eval($\phi$)} operator which accept temporal logic formula $\phi$ to be evaluated by means of a model checker tool. 
As a predictive quality indicator, \textbf{eval($\phi$)} can be monitored by specifying a Quality Constraint as we will see in the Smart Grid case study.  

\subsection{Performance Model}
In this work we focus our attention on KPIs which refers to quantifiable service performances, i.e. resource utilization, number of request served, etc. To better fit our case study, we select a M/M/1 queuing model to represent these type of indicators. The intuition is that such indicators represent resources whose arrival usage requests are determined by a Poisson process of parameter $\lambda$, whereas the resource service time follow an exponential distribution of parameter $\mu$.distributed arrives with a distribution usage is approximated queuing model by having where arrivals are determined by a Poisson process and job service times have an exponential distribution.

Let us assume a KPI as a variable $k$ whose values can range in set $V_K$ seen as:
\[
	k \in V_K = A_V \cup C_V \cup I_V
\]
where the subsets $A_V$, $C_V$ and $I_V$ have the following meaning:
\begin{description}
\item[$A_V$:] it is the set of \textit{Admissible Values} $k$ takes when the system is in a state which fulfills all the QCs defined on the KPI
\item[$C_V$:] it is the set of \textit{Critical Values}, i.e. limits/targets values, on which the system still meets the required quality but beyond which this is no longer true. 
\item[$I_V$:] it is the set of \textit{Inadmissible Values} $k$ takes when the system is in a state which does not fulfill at least a QCs defined on the KPI
\end{description} 

We assume that $V_K$ is totally ordered and its subsets are disjoint, that is:
\[ 
 \forall a,b,c : a\in A_V,b \in C_V, c \in I_V\; \text{s.t.}\; a < b < c 
\]

\begin{figure}[ht!]
\centering
\includegraphics[scale=0.4]{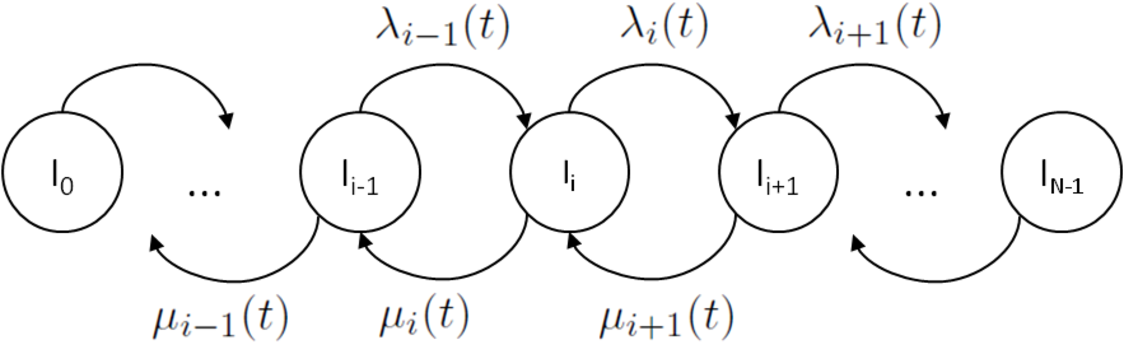}
\caption{The general form of the adopted queueing model}
\label{fig:queue}   
\end{figure}

The fig.\ref{fig:queue} illustrates our general queueing model. 
We consider a queue as a discrete representation of the set $V_K$. In particular, the $V_K$ is partitioned into a sequence of $N$ disjoint intervals ${I_i}=[a_i,b_i]$, $i\in{0,..,N-1}$ with $|I_i| = b_i-a_i = \frac{|V_K|}{N}$. Moreover, for $b_i \in I_i$, $a_{i+1}\in I_{i+1}$ for all $i \in {0,..,N-2}$, we have $b_i<a_{i+1}$. This helps to preserve the semantic distinction among the subsets $A_V$, $C_V$ and $I_V$. Hence, we can write $I_i < I_j$ if $i<j$. 

Thus, let $t$ be the total amount of elapsed time from the beginning of KPI monitoring, $w = t-T$ the time window, with $T<t$, in which we take into account the KPI variations,  and $k_{t_1}, k_{t_2}$, $t_1 < t_2$ two sequential values of the KPI of interest belonging respectively to the interval $I_i$ and $I_j$ with $i<j$.
We adapt the queueing model by interpreting:
\begin{itemize}
\item the queue length $L_Q = i$ as representing the interval $I_i$ in which $k_{t_1}$ lies
\item given all the transitions $I_i$ to $I_j$ with $i<j$ (resp. $i>j$) observed up to the time instant $t$, the increment (resp. decrement) rate $\lambda_t$ (resp. $\mu_t$) is $\sum_{i<j} \frac{j-i}{w}$ the ratio of the sum of all increments $j-i$ (resp. decrement $i-j$) over the time window we want to consider for the rate updating. 
\end{itemize} 

Therefore, the queue length increases from $L_Q=i$ to $L_Q=i+1$ for $i={0,..,N-1}$ with a rate $\lambda_t$ and decreases from $L_Q=i$ to $L_Q=i-1$ for $i={1,..,N}$ with a rate $\mu_t$.
An M/M/1 queue model can be described by an CTMC. In this way, by using the CSL as a language to formally specifying properties, we employ the probabilistic model-checking technique to conduct a quantitative analysis on the KPIs by means of their queue representation. 

\input{section4-casestudy}

\section{Conclusions and Future Work}
\label{sec:conc}
To support Big Data analysis of QoS information, in this paper, we have proposed a QoS prediction framework which takes advantage of the qualitative and quantitative analysis performed by a probabilistic model-checking technique.
Our approach uses a parametric QoS model and performs a probabilistic model-checking analysis in order to evaluate QoS-related predictive indicators (PIs). In this way, pre-alert QoS states can be notified in advance, giving a greater control to the Service Provider to avoid, or at least manage, possible breaches of Service Level Agreements (SLAs) contracted with Service Consumers.
We have realized and presented a validating prototype - built on top of the PRISM Model Checker, as well as experiments on a Smart Grid case study, which shows the effectiveness of our methodology and how, tuning the model parameters, the time required to model check is less than the time needed to receive updated QoS information from the monitored service. In the next future we plan to extend the experimental campaign validating our approach and to extend the usage of this framework to monitor security \cite{SEC} and other non-functional aspects other than provided QoS.

%% file: rw.tex
QoS prediction is surveyed in \cite{SCS+11,IKLL12,LWR+10,LMR+10}. A prediction performance model is treated in \cite{SCS+11}, where the authors exploit the Markovian Arrival Process (MAP) and a MAP/MAP/1 queuing model as a means to predict performance of servers deployed in Cloud infrastructure. 
Although in our Smart Grid case study we use a M/M/1 queuing model, our QoS prediction methodology does not rely on a 
specific model which, therefore, could be adapted as needed.
In \cite{IKLL12} is proposed a prediction-based resource measurement which use Neural Networks and Linear Regression as techniques to forecast future resource demands. A regression model is used also in \cite{LWR+10} to produces numerical estimation of Service Level Objectives (SLOs)
so as to predict SLA violations at runtime. Similarly, in \cite{LMR+10} is presented the PREvent framework, a system which uses a regression classifier to predict violation, but not details are given about the performance of the method.
In \cite{CCG+09,CBT13,YYSH08} the QoS requirements are controlled by solving a QoS optimization problem at runtime. Particularly, in \cite{CCG+09} a linear programming 
optimization problem is adopted to define a runtime adaptation methodology for meeting QoS requirements of service-oriented systems, whereas a 
multi-objetive optimization problem is proposed in \cite{CBT13,YYSH08} to develop QoS adaptive service-based systems to guarantee pre-define QoS attributes.

A collaborative method is proposed in \cite{ZZL12} in which performance of cloud components are predicted based on usage experiences.
Although this method could be appropriate for QoS indicators from the user perspective, is impractical in general case where QoS are business-oriented.

A QoS prediction by using a Model Checking solution is proposed in \cite{GGMT08,GMZ13}.
Gallotti et al. in \cite{GGMT08} propose an approach named ATOP - i.e. from Activity diagrams TO Prism models - 
which from an abstract description of service compositions (activity diagram) derives a probabilistic model to feed the PRISM tool for 
the evaluation phase. However, unlike our solution this is a methodology conceived for evaluating system at design-time. 
Similar to our work, in \cite{GMZ13} the authors propose a two-phase method involving monitoring and prediction with the aim of monitoring 
at run-time the reliability of compositional Web services which exhibit random behaviour. Although this method also takes advantage of the 
probabilistic model checking technique, it focuses mainly on reliability by providing a DTMC-based Markovian model. In contrast, we propose
a general CTMC probabilistic model for performance indicators in which both state and transition are parameterised, resulting in a model adaptable at run-time.

%% file: section4-casestudy.tex
\section{The Smart Grid Case Study}
\label{sec:casestudy}
The proposed QoS Prediction approach has been validated with respect to a Smart Grid (SG) case study. 

SG is the integration of the IT infrastructure into a traditional power grid in order to continuously exchange and process information to better control the production, consumption and distribution of electricity. For this purpose Smart Meters (SMs) devices are used to measure variations of electric parameters (e.g. voltage, power, etc.) and send such data to a computational environment which, in turn, analyse and monitor it in a real-time fashion.

\begin{figure*}[!htb]
\centering
\includegraphics[scale=0.6]{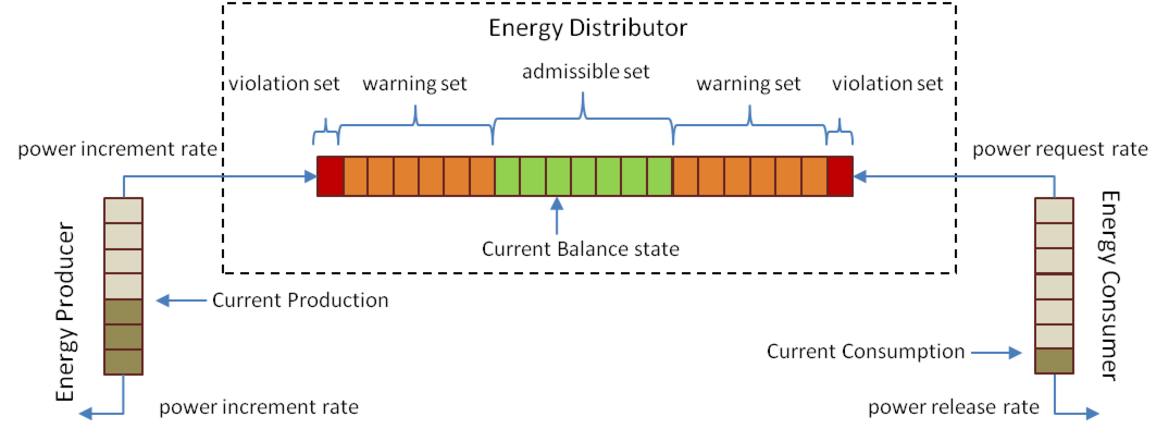}
\caption{Network queuing system model}
\label{fig:sg-model}    
\end{figure*}

In this case study, our tool performs the remote monitoring on behalf of an Energy Distributor (ED) which purchases electric power from Energy Producers (EPs) and retails it to Energy Consumers (ECs). The primary goal of the ED is to balance the purchased electric power with respect to the variations of power demand. 
\\
\textbf{The SG Model.}
For the sake of simplicity we have built a basic model which represents an ED, EP (or aggregated values of many EPs) and EC (or aggregated values of many ECs) as a three-queue system networked as in fig. \ref{fig:sg-model}. 
Each queue is a discrete representation of the real-valued KPI to be modelled. 
The PRISM-based model we define implements the queues $ED_Q$, $EP_Q$, and $EC_Q$ with queue length and transition rates as parameters. 
Furthermore, the sets $A_V$, $C_V$, $I_V$ are arranged as follows:
\[ \footnotesize
\begin{split}
	A_V &= \{d_e : d_e\ge adm_{min}\; \text{and}\; d_e\le adm_{max}\} \\
	C_V &= \{d_e : -cri_{min}\le d_e< adm_{min}\}\; \cup \\ 
		& \quad \cup \{d_e: adm_{max}< d_e \le cri_{max}\} \\
	I_V &= \{d_e : d_e<-cri_{min}\; \text{or}\; d_e>cri_{max}\}
\end{split}
\]
where $adm_{max}$, $adm_{min}$ and $cri_{min}$ and $cri_{max}$ represent the minimum and maximum thresholds of the admissible and critical value sets.
\\ 
\\
\textbf{Parameters Updating.}
In the queuing model the queue length and the transition rate are updated as follows.
Two thresholds are set on both queue edges so that if the current state goes below the first or up the second, the queue length is doubled or halved respectively. 
As for the updating of the transition rate, an Exponential Weighted Moving Average (EWMA) is applied on the first difference of the time series 
under analysis. Thus, let $Y = y_1,y_2,...$ a time series, we compute the transition rate $\rho$ as follows:
\begin{equation} \label{eq:update}
\rho' = \alpha(y_i - y_{i-1})+(1-\alpha)\rho
\end{equation}
in which the initial value of $\rho$ is set to $0$. The \ref{eq:update} is used for both the increment ($\mu_t$) and the decrement rate ($\lambda_t$).
\newline\newline
\textbf{QoS Data Extraction.}
To tackle the Big Data issue, our architecture takes advantage of a Complex Event Processing (CEP) which could be performed on any data-intensive distributed framework (e.g. Hadoop). Such combination allows to extract, process and deliver (complex) data in real-time, empowering the QoS monitoring and prediction phases.
Following our case study, we show an example of a complex event to derive the \textit{balance indicator} (BI) from the basic SmartMeterMeasureEvent originating from the Smart Meters of EPs and ECs: 
\begin{verbatim}
insert into BalanceIndicatorEvent
select (EP.measure - EC.measure) as index 
from EP.SmartMeterMeasureEvent as EP, 
	 EC.SmartMeterMeasureEvent as EC,

select "range_i"
from BalanceIndicatorEvent
where index index >= I_min and <= I_max
\end{verbatim}

The first query compute the BI and create the event \verb|BalanceIndicatorEvent|. The second query is a template used by our QoS Monitoring tool to generate the actual queries. They are used to classify at which range the index belongs to. Based on this data, the QoS Analyser component compute the transition rate from one range to another to be fed the PRISM model. 
On the other side, a temporal-based query is used for a real-time anomalous detection:
\begin{verbatim}
select measure, "CriticalValueMsg" 
from EP1.SmartMeterEvent.win:time(15 min)
where measure < BASE_PROD
\end{verbatim}

In this case we take advantage of the temporal-based capability of the CEP language. The select deliver a \verb|CriticalValueMsg| message based on the fact that a specific energy producer (EP1 in the example) is gone underproduction. The message is delivered to the QoS Monitoring which in turn perform the associated action, e.g. notify ED.
\newline\newline
\textbf{Quality Constraints to be Monitored.}
Briefly we report only two types of QCs: the first is a safety property (neither \textit{within} nor \textit{along} operator specified) which assesses if the predicted violation probability in the next 15 minutes is more than 10\%.
\begin{equation} \label{prop:1}
\footnotesize
	\text{\bf eval}(\mathcal{P}_{\ge 0.1}[F_{\le 30} \verb|"violState"|]) = \verb|false|
\end{equation}

The second QC guarantees to be notified if the probability of incurring in a violation state in the next 30 minutes is greater than 0.05 twice in a row (considering a measurement events every 15 minutes). 

\begin{equation} \label{prop:2}
\footnotesize
	\text{\bf eval}(\mathcal{P}_{=?}[F_{\le 30} \verb|"violState"|]) \le 0.05\; \text{\bf within}\; 30m\; 
\end{equation}

\subsection{Validation}
\label{sec:valid}
\begin{table*}[!htb] \footnotesize
\centering
\begin{tabular}{c|ll|lll}
\hline
\bf Queue length  & \bf \#States & \bf \#Trans & \bf BM time (s) & \bf MC time (s) & \bf Tot. time (s)\\ \hline 
 20 & 13280 & 63476 & 0.15 & 0.38 & 0.53 \\ 
 40 & 95360 & 466156 & 0.81 & 8.37 &  9.18\\ 
 60 & 310240 & 1528036 & 7.75 & 52.51 & 60.27 \\ 
 80 & 721920 & 3569116 & 20.53 & 197.11 & 217.64 \\ 
 100 & 1394400 & 6909396 & 42.80 & 492.28 & 535.089 \\ 
\hline 
\end{tabular} 
\caption{Queues length, model size, and execution time}
\label{tab:modelsize-time}
\end{table*}

In our scenario we assume a balance range of 800 Mega Watt (MW), i.e. $[min_b=-400,max_b=400]$, and we firstly evaluate how much time the QoS prediction phase takes with respect to different model size (Table \ref{tab:modelsize-time}). 
The table also reports the size of the model in terms of number of states and transitions. As expected by using a model-ckecking technique, the time is exponential against the model size. However, as the last row shows, we can also observe that even in case of millions ($10^6$) of states and transitions - that means a fine-grained discretisation - the total time is less than 9 minutes, hence still comparable with the updating rate usually considered for SGs. 

We have selected a queue length of 40 - i.e. a unit increment/decrement of the queue correspond to a 20MW of balance variation - and set these thresholds: $adm_{min} = -200$, $adm_{max} = 200$, $cri_{min} = -380$, $cri_{max} = 380$. Our tests are based on property \ref{prop:2} evaluated by simulating three different scenarios: 
\begin{description}
\item[Case A:] EPs inject in the grid as much energy as ECs need (balanced case).\\
\item[Case B:] ECs request less than EPs produce (overproduction).\\
\item[Case C:] The energy consumption request rises twice compared with the production rate (imbalanced condition).
\end{description}

\begin{figure}[!htbp]
\centering
\includegraphics[scale=0.4]{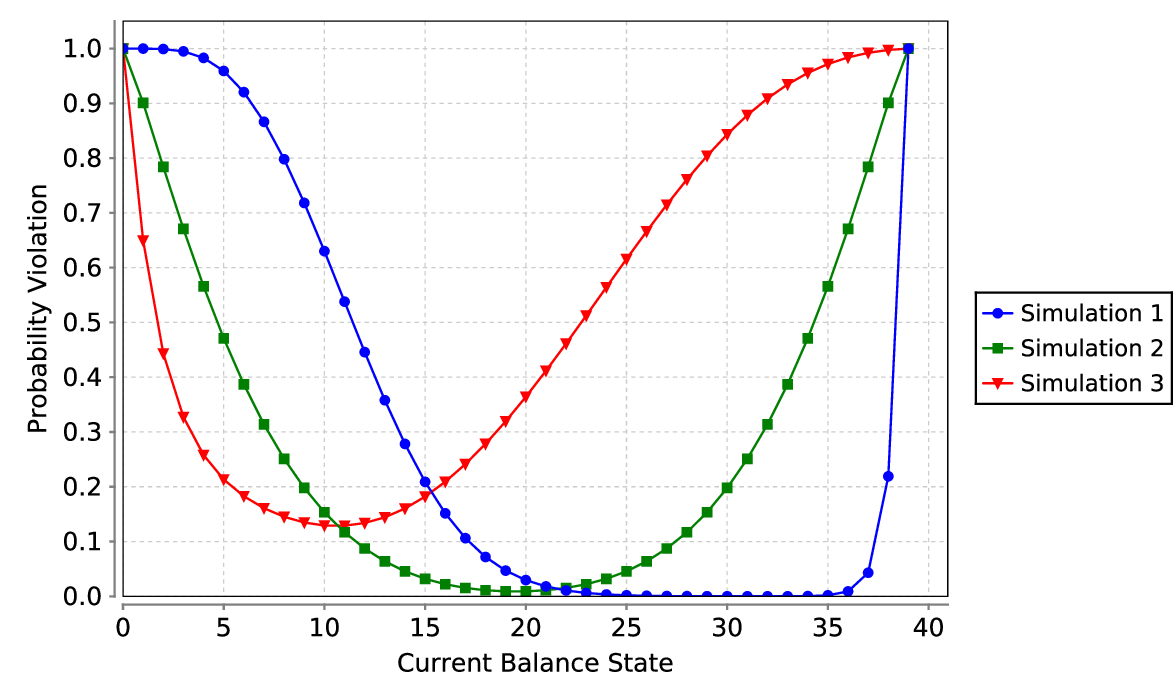}
\caption{Violation Probability (queue length=40)}
\label{fig:viol-balance}     
\end{figure}
Fig. \ref{fig:viol-balance} plots the violation probability estimated for such scenarios.
For scenario A the violation probability varies in a symmetrical fashion around the balance point (i.e. queue length 20).  
The scenario B exhibits a higher probability in all the overproduction states (i.e. queue length less than the balance point), and a lower one for a large number of states representing the power grid overload (i.e. queue length greater than the balance point). 
This characteristic is emphasised in the third simulation which represents the imbalanced (overloaded in this case) scenario. In addition, we notice how in such anomalous conditions all minimum values of violation probability are higher than the other two scenarios.

%% file: references.tex
%
%
%
%
